\documentclass[12pt]{article}

\newsavebox{\ns}
\newsavebox{\dbrane}
\newsavebox{\dbshort}

\usepackage{epsfig}
\usepackage{amssymb}

\def\appendix{{\newpage\section*{Appendix}}\let\appendix\section%
        {\setcounter{section}{0}
        \gdef\thesection{\Alph{section}}}\section}

\def\be{\begin{eqnarray}}
\def\ee{\end{eqnarray}}

\newcommand{\nn}{\nonumber}

\newcommand{\ft}[2]{{\textstyle\frac{#1}{#2}}}
\newcommand{\eqn}[1]{(\ref{#1})}

\newcommand\bomega{\mbox{\boldmath $\omega$}}

\newcommand\bmu{\mbox{\boldmath $\mu$}}

\newcommand\btau{\mbox{\boldmath $\tau$}}

\def\Dslash{\,\,{\raise.15ex\hbox{/}\mkern-12mu D}}
\def\Dbarslash{\,\,{\raise.15ex\hbox{/}\mkern-12mu {\bar D}}}
\def\delslash{\,\,{\raise.15ex\hbox{/}\mkern-9mu \partial}}
\def\delbarslash{\,\,{\raise.15ex\hbox{/}\mkern-9mu {\bar\partial}}}
\def\pslash{\,\,{\raise.15ex\hbox{/}\mkern-9mu p}}
\def\calDslash{\,\,{\raise.15ex\hbox{/}\mkern-12mu {\cal D}}}

\begin{document}
\pagestyle{plain}
\setcounter{page}{1}
\newcounter{bean}
\baselineskip16pt

\begin{titlepage}

\begin{center}
\today
\hfill hep-th/0204186\\
\hfill MIT-CTP-3266 \\

\vskip 1.5 cm
{\large \bf NS5-Branes, T-Duality and Worldsheet Instantons}
\vskip 1 cm 
{David Tong}\\
\vskip 1cm
{\sl Center for Theoretical Physics, 
Massachusetts Institute of Technology, \\ Cambridge, MA 02139, U.S.A.\\}

\end{center}

\vskip 0.5 cm
\begin{abstract}
The equivalence of NS5-branes and ALF spaces under T-duality 
is well known. However, a naive application of T-duality  
transforms the ALF space into a smeared NS5-brane, de-localized 
on the dual, transverse, circle. In this paper we re-examine this 
duality, starting from a two-dimensional ${\cal N}=(4,4)$ gauged 
linear sigma model describing Taub-NUT space. After dualizing the 
${\bf S}^1$ fiber, we find that the smeared NS5-brane target space 
metric receives corrections from multi-worldsheet instantons. These 
instantons are identified as Nielsen-Olesen vortices. We show that their 
effect is to break the isometry of the target space, localizing 
the NS5-brane at a point. The contribution from the $k$-instanton 
sector is shown to be proportional to the weighted integral of the 
Euler form over the $k$-vortex moduli space. The duality also predicts 
the, previously unknown, asymptotic exponential decay coefficient of 
the BPS vortex solution.

\end{abstract}

\end{titlepage}

\section{Introduction and Summary}

Under T-duality of type II string theory, NS5-branes are mapped 
into Ricci flat, background geometries. For $N$ parallel NS5-branes, 
this background is the hyperK\"ahler metric on an asymptotically 
locally flat (ALF) space with an $A_{N-1}$ singularity \cite{ov}. 
This relationship plays a prominent role in the duality plexus yet, 
from the perspective of the string worldsheet, has 
been proven only for {\it smeared} NS5-branes, for which the 
transverse circle is an isometry and the usual Buscher rules for 
T-duality may be applied \cite{buscher,tduality}. This is an 
unsatisfactory state of affairs. The purpose of this paper is to 
rectify this situation and demonstrate T-duality between the 
ALF space and the {\it localized} NS5-brane. As we shall see, 
the missing ingredient is the contribution from worldsheet instantons.

Our tool in exploring this duality is the ${\cal N}=(4,4)$ 
supersymmetric gauged linear sigma model for 
the ALF space. Since their inception, linear sigma models 
have proven useful in determining the effects of worldsheet 
instantons \cite{witten,mp}. More pertinently, Hori and Vafa have 
recently used this technique to calculate instanton corrections to 
T-duality transformations in theories with ${\cal N}=(2,2)$ 
supersymmetry \cite{hv}. Subsequent applications include  
\cite{hk}. In each of these cases, a superpotential is generated 
by a one instanton effect, breaking a global symmetry of the theory. 
While the calculations presented below are similar to those of 
\cite{hv,hk}, they differ in two important respects. 
Firstly, the existence of ${\cal N}=(4,4)$ supersymmetry prohibits 
the generation of a superpotential, and the instantons now correct 
the target space metric where their effect is to break an isometry. 
Secondly, the corrections do not stop at the one instanton level, and  
the metric receives contributions from all topological sectors.

To make this discussion more explicit, let us start with type II 
string theory ``compactified'' on an ALF space. For simplicity 
we consider Taub-NUT space. It is worth noting that this is 
T-dual to a single NS5-brane, and subtleties abound concerning 
this object \cite{chs}. However, these difficulties play no role 
in the following discussion. 
The metric of Taub-NUT space is given by,
\be
ds_{ALF}^2=H(r)d{\bf r}\cdot
d{\bf r}+\frac{1}{4}H(r)^{-1}
(d\psi+\bomega\cdot d{\bf r})^2
\label{tn}\ee
where ${\bf r}\in {\bf R}^3$ and $\psi\in [0,4\pi)$ and 
$\nabla\times\bomega =\nabla(1/r)$. The harmonic function $H(r)$ is 
given by,
\be
H(r)=\frac{1}{g^2}+\frac{1}{2r}
\label{hsmear}\ee
As $r\rightarrow 0$, the metric becomes flat ${\bf R}^4$ while, 
in the asymptotic regime $r\rightarrow\infty$, the metric 
is locally ${\bf R}^3\times {\bf S}^1$, with the ${\bf S}^1$ 
parameterized by $\psi$. The asymptotic radius of this circle is $g$. 

This metric enjoys a $U(1)$ isometry 
which acts by shifting the value of $\psi$, ensuring that momentum 
around the circle is a conserved quantum number. In contrast, string 
winding number around the circle is not conserved since the string 
may slip off, either by moving to $r=0$ where the circle degenerates, 
or alternatively by wrapping once around the asymptotic boundary of 
Taub-NUT \cite{ghm}. 

The existence of the $U(1)$ isometry allows for the application of the 
usual T-duality transformation, exchanging momentum and winding on the 
string worldsheet: 
$\partial_\mu\psi = g^2\epsilon_{\mu\nu}\partial^\nu\theta$. 
As we will review in Section 2, after such an operation the resulting metric 
has the form,
\be
ds_{NS5}^2=H(r)\left(d{\bf r}\cdot d{\bf r}+d\theta^2\right)
\label{metric}\ee
where $\theta\in [0,2\pi)$ parameterizes the dual ${\bf S}^1$. The 
conformal factor $H$ is related to the dilaton and 
is given by \eqn{hsmear}, implying that the dual circle has 
asymptotic radius $1/g$, as 
expected under T-duality. Moreover, as we shall see explicitly in 
Section 2, the 
non-trivial fibration of ${\bf S}^1$ over ${\bf R}^3$ in 
Taub-NUT space results in a torsion term, 
$T_{ijk}=\epsilon_{ijk}^{\ \ \ l}\partial_lH^{-1}$, carrying the charge 
of a single NS5-brane. However, as may be seen from 
\eqn{hsmear} and \eqn{metric}, the metric has no dependence on the dual 
circle $\theta$. In other words, the NS5-brane has been smeared in 
this direction.

Thus, after the naive application of T-duality, we find a string 
background in which both momentum and winding are conserved. 
Let us contrast this situation with the metric for an NS5-brane that 
is fully 
localized on ${\bf R}^3\times{\bf S}^1$. This can be easily 
determined from supergravity by considering an infinite, periodic, 
array of colinear NS5-branes. Upon  
Poisson resummation, the metric takes the form \eqn{metric},  
but with the smeared function \eqn{hsmear} replaced by \cite{ghm}
\be
H_{sugra}(r,\theta)=\frac{1}{g^2}+\frac{1}{2r}\left(
\frac{\sinh r}{\cosh r-\cos \theta}\right)
\label{h}\ee
Notice in particular the appearance of $\theta$ in the function 
$H_{sugra}$, ensuring that the circle transverse to the NS5-brane is no 
longer an isometry\footnote{To compare with the metric for the NS5-brane in 
flat space, define $\hat{\bf r}={\bf r}/t$ and 
$\hat{\theta}=\theta /t$, and send $t\rightarrow 0$, keeping 
$\hat{{\bf r}}$ and $\hat{\theta}$ fixed.}, and momentum in this direction
is no longer conserved. 

Our challenge is to reproduce this localization of 
the NS5-brane from the worldsheet perspective of T-duality. A clue as to 
the mechanism responsible for this localization can be found from 
the Taylor expansion \cite{ghm},
\be
H_{sugra}(r,\theta)=\frac{1}{g^2}+\frac{1}{2r}\left(1+
\sum_{k=1}^\infty\sum_\pm e^{-kr\pm ik\theta}\right)
\label{exp}\ee
which points the finger of responsibility firmly at worldsheet 
instantons. Thus, in Section 3 we turn to the task of uncovering 
the instanton structure of the gauge theory. As we 
shall see, although the theory does contain the topology necessary to 
admit instantons, there are in fact no finite action 
solutions to the equations of motion. The situation is 
reminiscent of constrained instantons in four-dimensional 
Yang-Mills-Higgs theories \cite{affleck}, and we proceed 
accordingly. The lessons of four dimensional instanton 
calculations teach us that exact results may be gleaned from 
constrained instantons, providing the approximate solutions 
are suitably compatible with supersymmetry (see for example  
\cite{dkmadhm}). 
We therefore identify approximate Bogomoln'yi type solutions to 
the equations of motion. These 
solutions are nothing more than the BPS vortices of the abelian 
Higgs model. In the remainder of Section 3 we compute the 
various components necessary to perform the instanton 
computation and piece everything together to 
find that the isometry of the smeared NS5-brane is indeed broken, with 
the instanton corrections to the smeared function $H(r)$ taking the form,
\be
H_{inst}(r,\theta)=\frac{1}{g^2}+\frac{1}{2r}\left(1+
\sum_{k=1}^\infty\sum_\pm m_k e^{- kr\pm ik\theta}\right)
\label{hinst}\ee
where the numerical coefficients $m_k$ encode 
certain information about the classical BPS vortex solutions. Due to 
the lack of integrability of the vortex equations, the results necessary 
to determine $m_k$ from first principles are not available in the 
soliton literature. For example, neither the one-vortex solution, 
nor the two-vortex moduli space is known analytically, facts which 
impede calculation of $m_1$ and $m_2$ respectively. 
However, we may use the conjectured equality,
\be
H_{inst}(r,\theta)=H_{sugra}(r,\theta)
\nn\ee
to yield predictions for these quantities; namely  
$m_k=1$ for all $k$. For the $k=1$ instanton sector, we will 
show that $m_1=l_1^4/8$, where the 
coefficient $l_1$ characterizes the exponential radial decay of 
the Higgs field in the single vortex solution 
(see equation \eqn{qlarge}). This number is known only through 
numerical studies \cite{vega,speight}. However, agreement 
with supergravity predicts $l_1=8^{1/4}$. It is heartening 
that this value agrees with \cite{speight} to within one 
part in a thousand (and is within 
$2\%$ of the older numerical value quoted in \cite{vega}).

For the higher instanton sectors, the numbers $m_k$ do not 
have as simple an interpretation. They are given by
\be
m_k =  (2k)^{-3}\, \nu(\tilde{\cal M}_k)
\ee
where $\tilde{\cal M}_k$ is the centered moduli space describing the 
relative positions of $k$ BPS vortices in the abelian Higgs model, 
and $\nu(\tilde{\cal M}_k)$ is the weighted integral of the Euler 
form over ${\cal M}_k$ (see equation \eqn{arnold}). 
The appearance of the Euler form for 
different soliton moduli spaces is not uncommon in instanton 
calculations. In 
particular, it was first encountered by Dorey, Khoze and Mattis in 
the context of three-dimensional gauge theories \cite{dkm}, and 
the calculation presented here closely follows this paper. One 
difference in the present case is that this integral is weighted 
by the functions $l_k^4$, which characterize the exponential radial 
decay of the $k$-vortex solution. It would be interesting to understand 
these functions in more detail.

Other approaches relating the NS5-brane to ALF spaces include 
an analysis of the 5-brane worldvolume dynamics \cite{berg}, 
mirror symmetry 
in three-dimensional gauge theories \cite{ds}, Nahm transforms 
\cite{kraan} and D-brane probes \cite{karch}.

%A particularly intriguing approach involves 
%examining mirror pairs of three-dimensional gauge theories 
%\cite{is} compactified on ${\bf R}^2\times {\bf S}^1$ 
%\cite{ds}. Recall that mirror symmetry in three-dimensional gauge 
%theories interchanges the Higgs and Coulomb branches of the two 
%theories. Below the compactification scale, the theory on the 
%Higgs branch reduces to a non-linear sigma-model on the vacuum 
%moduli space. In contrast, in the dual theory on the Coulomb 
%branch, one must integrate out an infinite tower of Kaluza-Klein 
%modes, which correct the metric and generate a torsion term 
%\cite{ds}. In this fashion one can reproduce T-duality of 
%two-dimensional conformal theories from mirror symmetry of 
%three-dimensional theories. Further aspects of this correspondence, 
%including the application to theories with less supersymmetry, have 
%been explored in \cite{ahkt}.
%{\bf put at end?}

\section{Gauge Theory, ALF Spaces and T-Duality}

In this section we introduce the gauge theory of interest. 
We start by describing the full supersymmetric gauge 
theory in all its glory. We show that in the classical, 
low-energy limit, this gauge theory reduces to the sigma-model 
with torsion describing the NS5-brane smeared on a transverse 
circle, as discussed in the introduction. In Section 2.2 
we perform a T-duality transformation, and show that this result 
is equivalent to the Taub-NUT metric. Appendix A contains a discussion 
of the generalization to $A_{N-1}$ ALF spaces.

\subsection{The Gauge Theory}

Let us start with a description of the gauge theory. It has 
${\cal N}=(4,4)$ supersymmetry (or 8 non-chiral supercharges) 
in $d=1+1$ dimensions, and may be thought of as the 
${\cal N}=(4,4)$ extension of the theories discussed recently 
by Hori and Kapustin \cite{hk}. There are three superfield 
representations of the 
${\cal N}=(4,4)$ supersymmetry algebra: a vector multiplet, a 
hypermultiplet and a twisted hypermultiplet. To construct our 
gauge theory, we need one of each.

The ${\cal N}=(4,4)$ vector multiplet contains a $U(1)$ gauge 
field, two complex scalars $\phi$ and $\sigma$, and two Dirac 
fermions $\lambda$ and $\tilde\lambda$. It will prove convenient 
to decompose our ${\cal N}=(4,4)$ superfields into ${\cal N}=(2,2)$ 
representations, resulting in an ${\cal N}=(2,2)$ $U(1)$ vector multiplet 
$V$ and a neutral chiral multiplet $\Phi$. 
For the purposes of writing the gauge kinetic terms, one usually 
exchanges the vector multiplet in favor of a twisted chiral multiplet  
$\Sigma = \bar{D}_+ D_-V$. (For a detailed introduction to 
${\cal N}=(2,2)$ theories and superspace notation, see 
\cite{witten,hv}). The non-derivative terms in the component 
expansion of these superfields are,
\be
\mbox{Vector multiplet}\ \ \left\{ \begin{array}{l}
\Sigma=\sigma-i\sqrt{2}\theta^+\bar{\lambda}_+-i\sqrt{2}
\bar{\theta}^-\bar{\lambda}_-+\sqrt{2}\theta^+\bar{\theta}^-
(D^3-iF_{12})+\ldots
\nn\\
\Phi=\phi+\sqrt{2}\theta^+\tilde{\lambda}_++\sqrt{2}\theta^-
\tilde{\lambda}_-+\frac{1}{\sqrt{2}}\theta^+{\theta}^-
(D^1+iD^2)+\ldots\end{array}\right.
\nn\ee
The hypermultiplet contains two complex scalars, $q$ and $\tilde{q}$, which 
are again paired with two Dirac fermions, $\psi$ and $\tilde{\psi}$. 
It decomposes into two chiral multiplets $Q$ and 
$\tilde{Q}$ with charge $+1$ and $-1$ respectively under the $U(1)$ 
gauge group. Again, the lowest components in the expansion read,
\be
\mbox{Hypermultiplet}\ \ \ \left\{ \begin{array}{l}
Q=q+\sqrt{2}\theta^+{\psi}_++\sqrt{2}
{\theta}^-\psi_-+\theta^+\theta^-F+\ldots
\nn\\
\tilde{Q}=\tilde{q}+\sqrt{2}\theta^+\tilde{\psi}_++\sqrt{2}\theta^-
\tilde{\psi}_-+\theta^+{\theta}^-\tilde{F}+\ldots\end{array}\right.
\nn\ee
Finally, the twisted hypermultiplet also contains four scalars and two 
Dirac fermions. This time the scalars naturally pair off into a triplet 
${\bf r}=(r^1,r^2,r^3)$ and a singlet 
$\theta$. We denote the fermions as $\chi$ and 
$\tilde{\chi}$. Under decomposition into ${\cal N}=(2,2)$ 
superfields, we have a chiral 
multiplet $\Phi$ and a twisted chiral multiplet $\Theta$, each 
of which is uncharged under the gauge group. Once more, 
the component fields are given by,
\be
\mbox{Twisted Hypermultiplet}\ \left\{ \begin{array}{l}
\Psi=(r^1+ir^2)+\sqrt{2}\theta^+\bar{\chi}_++\sqrt{2}
{\theta}^-\chi_-+\theta^+\theta^-G+\ldots
\nn\\
\Theta=(r^3+i\theta)-i\sqrt{2}\theta^+\bar{\tilde{\chi}}_+
-i\sqrt{2}\bar{\theta}^-\tilde{\chi}_-+\theta^+\bar{\theta}^-
\tilde{G}+\ldots\end{array}\right.
\nn\ee
With these conventions, the action takes a simple form in 
superspace notation. The kinetic terms for all fields arise 
from D-terms,
\be
{\cal L}_D=\int d^4\theta\ \frac{1}{e^2} \left( \Sigma^\dagger \Sigma 
+\Phi^\dagger \Phi\right) + \frac{1}{g^2}\left(\ \Theta^\dagger\Theta + 
\Psi^\dagger\Psi\right) + Q^\dagger e^{2V}Q + \tilde{Q}^\dagger e^{-2V} 
\tilde{Q} 
\nn\ee
while the F-terms and twisted F-terms are given by,
\be
{\cal L}_F=\int d\theta^+d\theta^-\ \left(\tilde{Q} \Phi Q - \Psi\Phi
\right)\ \ \ \ ,\ \ \ \ 
{\cal L}_{\tilde{F}}= -\int d\theta^+d\bar{\theta}^-\ \Sigma\Theta
\nn\ee
The cubic superpotential is familiar from all theories with 8 
supercharges. The other two terms give the coupling between the 
vector multiplet and twisted hypermultiplet. 
These ensure that the fields 
${\bf r}$ play the role of a triplet of dynamical Fayet-Iliopoulos
(FI) parameters, while $\theta$ is a dynamical theta-angle. 

In component form, the action splits neatly into kinetic, scalar 
potential and fermion mass and 
Yukawa terms, 
\be
S=\frac{1}{2\pi}\int d^2x\ {\cal L}_{kin}+{\cal L}_{pot}+{\cal L}_{yuk}
\label{fullaction}\ee
where the pre-factor of $1/2\pi$ is the usual normalization of the 
Polyakov action, and ensures that T-duality acts as 
$(\mbox{radius})\rightarrow(\mbox{radius})^{-1}$, with no further 
numerical factors. We have,
\be{\cal L}_{kin}&=&\frac{1}{e^2}\left(\ft12F_{01}^2-
\ft12|\partial\phi|^2+\ft12|\partial\sigma|^2+i
(\bar{\lambda}_+\partial_-\lambda_++\bar{\tilde{\lambda}}_+
\partial_-\tilde{\lambda}_++\bar{\lambda}_-\partial_+\lambda_-
+\bar{\tilde{\lambda}}_-
\partial_+\tilde{\lambda}_-)\right) \nn\\
&&+\frac{1}{g^2}\left(-\ft12|\partial{\bf r}|^2-\ft12(\partial\theta)^2
+i(\bar{\chi}_+\partial_-\chi_++\bar{\tilde{\chi}}_+
\partial_-\tilde{\chi}_++\bar{\chi}_-\partial_+\chi_-+\bar{\tilde{\chi}}_-
\partial_+\tilde{\chi}_-)\right) \nn\\ 
&& +\left(-|{\cal D}q|^2-|{\cal D}\tilde{q}|^2+
i(\bar{\psi}_+{\cal D}_-\psi_++\bar{\tilde{\psi}}_+
{\cal D}_-\tilde{\psi}_++\bar{\psi}_-{\cal D}_+\psi_-+\bar{\tilde{\psi}}_-
{\cal D}_+\tilde{\psi}_-)\right)
\nn\ee
where our conventions are $\partial_\pm=\partial_0\pm\partial_1$, and 
${\cal D}q=\partial q-iAq$, and 
${\cal D}\tilde{q}=\partial\tilde q+iA\tilde q$. The scalar potential 
is given by,
\be
{\cal L}_{pot}&=&-\frac{e^2}{2}\left(|q|^2-|\tilde{q}|^2-r^3\right)^2
-\frac{e^2}{2}\left|2\tilde{q}q-(r^1+ir^2)\right|^2 + \theta F_{01} 
\nn\\ && -(|\phi|^2+|\sigma|^2)(|q|^2+|\tilde{q}|^2+g^2)
\nn\ee
The terms on the first line make manifest the roles played by the 
fields ${\bf r}$ and $\theta$ as FI parameters and theta angles 
respectively. In particular, it is clear from the 
topological nature of the $\theta F$ coupling that the physics is 
invariant under shifts of $\theta\rightarrow \theta + 2\pi$. 
It is worth noting that the supersymmetric 
completion of these couplings includes the mass $g$ of the vector 
multiplet scalar fields, which may be seen in the second line. 
Finally, the fermion masses and Yukawa couplings are given by,
\be
{\cal L}_{yuk}&=&-\left(\chi_-\tilde{\lambda}_++\tilde{\lambda}_-
\chi_++\bar{\tilde{\chi}}_+\lambda_-+\bar{\lambda}_+\tilde{\chi}_-
\right)-\left(\bar{\tilde{\lambda}}_+\bar{\chi}_-+\bar{\chi}_+
\bar{\tilde{\lambda}}_-+\bar{\lambda}_-\tilde{\chi}_++
\bar{\tilde{\chi}}_-\lambda_+\right) \nn\\
&&+i\sqrt{2}q\left(\bar{\lambda}_-\bar{\psi}_+-\bar{\lambda}_+\bar{\psi}_-
+i\tilde{\lambda}_-\tilde{\psi}_++i\tilde{\psi}_-\tilde{\lambda}_+
\right) -\sqrt{2}\sigma\left(\bar{\psi}_-\psi_+-\bar{\tilde{\psi}}_-
\tilde{\psi}_+\right) \nn\\
&& -i\sqrt{2}q^\dagger\left(\psi_+\lambda_--\psi_-\lambda_+
-i\bar{\tilde{\psi}}_+\bar{\tilde{\lambda}}_--i\bar{\tilde{\lambda}}_+
\bar{\tilde{\psi}}_-\right) -\sqrt{2}{\sigma}^\dagger\left(\bar{\psi}_+\psi_-
-\bar{\tilde{\psi}}_+\tilde{\psi}_-\right)\nn\\
&& +i\sqrt{2}\tilde{q}\left(\bar{\lambda}_-\bar{\tilde{\psi}}_+-
\bar{\lambda}_+\bar{\tilde{\psi}}_--i\tilde{\lambda}_-\psi_++i\psi_-
\tilde{\lambda}_+\right) -\sqrt{2}\phi\left(\tilde{\psi}_-\psi_+
-\tilde{\psi}_+\psi_-\right)\nn\\
&&-i\sqrt{2}\tilde{q}^\dagger\left(\tilde{\psi}_+\lambda_-
-\tilde{\psi}_-\lambda_++i\bar{\psi}_+\bar{\tilde{\lambda}}_-
-i\bar{\tilde{\lambda}}_+\bar{\psi}_-\right)-\sqrt{2}{\phi}^\dagger
\left(\bar{\psi}_+\bar{\tilde{\psi}}_--\bar{\psi}_-
\bar{\tilde{\psi}}_+\right)
\nn\ee
The theory enjoys an $SU(2)_R\times SO(4)_R$ R-symmetry group, under 
which the vector multiplet scalars transform in the $({\bf 1},{\bf 4})$, 
the hypermultiplet scalars transform in the $({\bf 2},{\bf 1})$, and the 
twisted hypermultiplet scalars transform as $({\bf 1}+{\bf 3},{\bf 1})$. 
We have chosen to normalize fields such that the engineering 
dimension of the gauge coupling constant is $[e^2]=2$, which 
implies the vector multiplet scalars have $[\phi]=[\sigma]=1$, 
while all other scalar fields are dimensionless, $[q]=[\tilde{q}]=
[{\bf r}]=[\theta]=0$. Importantly, the coefficient of the 
twisted hypermultiplet kinetic terms is also dimensionless: $[g^2]=0$.

\subsubsection*{\rm{\it The Low-Energy Theory}}

The vacuum moduli space of the theory, determined by the zero locus 
of the scalar potential, is given by
\be
F_{01}=\sigma=\phi=0\ \ ,\ \ |q|^2-|\tilde{q}|^2=r^3\ \ ,\ \ 
2\tilde{q}q=r^1+ir^2
%\omega^\dagger\btau\omega={\bf r}\ \ \ 
\label{vacuum}\ee
Since $e^2$ is 
the only dimensionful parameter in the theory, low energies 
corresponds to $e^2\rightarrow\infty$, and the physics is 
described by a sigma-model on the vacuum moduli space \eqn{vacuum}. 
The metric on the moduli space is inherited from the kinetic 
terms which, in this case, receive contributions from both 
the hypermultiplet and twisted hypermultiplet. To see this, it 
is useful to introduce the gauge-variant field $\alpha=-2\arg (iq)$. 
After imposing the constraint \eqn{vacuum}, the hypermultiplet 
bosonic kinetic terms may then be written as,
\be
|{\cal D}q|^2+|{\cal D}\tilde{q}|^2 = \frac{1}{4r}\partial{\bf r}\cdot 
\partial{\bf r}+\frac{r}{4}(\partial\alpha+2A+{\bomega}\cdot 
\partial{\bf r})^2
\label{kin}\ee
where $\nabla\times\bomega=\nabla(1/r)$, and we have used the fact 
that, in vacuo, $r=|{\bf r}|=|q|^2+|\tilde{q}|^2$. We now chose to work in 
$\alpha=0$ gauge, and integrate out the gauge field $A$. In the strict 
$e^2\rightarrow\infty$ limit, the 
kinetic terms for $A$ vanish, but the $\theta dA$ term does not. 
We have,
\be
A_\mu = -\ft12 \bomega\cdot\partial_\mu {\bf r}+
\frac{1}{2r}\epsilon_{\mu\nu}
\partial^\nu\theta
\nn\ee
After substitution into the kinetic terms \eqn{kin}, and including 
the twisted hypermultiplet kinetic terms, we find that 
the bosonic sector of the low-energy effective action is given by 
a sigma-model with torsion,
\be
{\cal L}_{bosonic}=\frac{1}{2}H(r)\left(\partial_\mu{\bf r}
\partial^\mu{\bf r}+
\partial_\mu\theta\partial^\mu\theta\right)+\frac{1}{2}
\epsilon_{\mu\nu}\bomega\cdot\partial^\mu{\bf r}\,\partial^\nu\theta
\label{henry}\ee
where the function $H(r)$ is given by,
\be
H(r)=\frac{1}{g^2}+\frac{1}{2r}
\nn\ee
As discussed in the introduction, this is the target space metric and 
torsion describing an NS5-brane localized in the ${\bf R}^3$ 
parameterized by ${\bf r}$, but smeared along a transverse circle 
parameterized by $\theta$. In the following section, we shall 
see how quantum effects alter this metric. However, let us 
first continue our examination of the classical low-energy 
effective action, turning now to the fermions. Working once 
more in the infra-red limit $e^2\rightarrow\infty$, we notice 
that the vector multiplet fermions, $\lambda$ and $\tilde{\lambda}$, 
become Grassmannian Lagrange multipliers, imposing the fermionic 
constraints,
%\be
%\chi_+=\sqrt{2}\tilde{q}\psi_+-\sqrt{2}q\tilde{\psi}_+\ \ \ &,&\ \ \ 
%\bar{\tilde{\chi}}_+=-i\sqrt{2}\tilde{q}^\dagger\tilde{\psi}_+-i
%\sqrt{2}q^\dagger\psi_+
%\nn \\
%\chi_-=-\sqrt{2}\tilde{q}\psi_--\sqrt{2}q\tilde{\psi}_-
%\ \ \ &,&\ \ \ \bar{\tilde{\chi}}_-
%=i\sqrt{2}\tilde{q}^\dagger\tilde{\psi}_--i\sqrt{2}q^\dagger\psi_-
%\nn\ee
\be
\psi_+=\frac{1}{\sqrt{2}r}(iq\bar{\tilde{\chi}}_++\tilde{q}^\dagger
\chi_+)\ \ \ &,&\ \ \ \tilde{\psi}_+=\frac{1}{\sqrt{2}r}
(-q^\dagger\chi_++i\tilde{q}\bar{\tilde{\chi}}_+)\nn\\
\psi_-=\frac{1}{\sqrt{2}r}(iq\bar{\tilde{\chi}}_--\tilde{q}^\dagger
\chi_-)\ \ \ &,&\ \ \ \tilde{\psi}_-=\frac{1}{\sqrt{2}r}
(-q^\dagger\chi_--i\tilde{q}\bar{\tilde{\chi}}_-)
\label{chipsi}\ee
Substituting these into the fermionic kinetic terms, and summing 
the contribution from both the hypermultiplet and twisted hypermultiplet, 
we have
\be
{\cal L}_{fermionic}=iH(r)\left(\bar{\chi}_+\partial_-\chi_+
+\bar{\chi}_-\partial_+\chi_-+\bar{\tilde{\chi}}_+\partial_-
\tilde{\chi}_++\bar{\tilde{\chi}}_-\partial_+\tilde{\chi}_-\right)
+{\cal O}(\chi^2\partial q)
\nn\ee
This combines with the bosonic action \eqn{henry} to yield a 
the first two terms of an ${\cal N}=(4,4)$ supersymmetric sigma-model 
with torsion. The full supersymmetric action for this theory was 
given in \cite{sigmatorsion} and a nice summary may be found in 
\cite{braden}. The full action takes the form,
\be
S_{susy}&=&\frac{1}{2\pi}\int d^2x\ 
\ft12 G_{ij}\partial_\mu X^i\partial_\nu X^j 
+\ft12 B_{ij}\epsilon_{\mu\nu}\partial^\mu X^i\partial^\nu X^j 
+iG_{ij}\bar{\Omega}^i D\Omega^j \nn\\ &&\ \ \ \ \ \ \ \ \ \ \ \ \ \ 
+\ \ft14R_{ijkl}\bar{\Omega}^i
\Omega^j\bar{\Omega}^k\Omega^l
\label{compare}\ee
where the covariant derivative is given by 
$D\Omega^j=\partial\Omega^j+\Omega^i\Gamma_{ik}^j\partial X^k$. The 
connection $\Gamma$ in this expression is constructed with respect to 
the torsion $T=dB$, and the Riemann tensor $R$ is similarly defined. 
Comparing to the terms above, we have the metric $G$ given by 
equation \eqn{metric}, and the fields normalized as 
$X^i=(r^1,r^2,r^3,\theta)$ and $\Omega^i=(\chi_+,\chi_-,\tilde{\chi}_+,
\tilde{\chi}_-)$. 

\subsection{T-Duality and Taub-NUT Space}

In this section, we would like to perform T-duality on the 
periodic direction $\theta$, and show that the low-energy 
physics is determined by a sigma-model without torsion, with the 
target space metric given by Taub-NUT of equation \eqn{tn}. We work 
with the full 
gauge theory \eqn{fullaction} rather than the low-energy fields. 
To perform the duality, we first isolate the terms involving $\theta$, 
and then introduce the auxiliary Lagrangian,
\be
{\cal L}_{dual}=\frac{1}{2g^2}C_\mu C^\mu -\epsilon_{\mu\nu}
C^\mu A^\nu +\epsilon_{\mu\nu}\partial^\mu C^\nu \kappa 
\nn\ee
where the topological nature of the final term ensures that physics 
is invariant under $2\pi$ shifts of the Lagrange multiplier $\kappa$. 
Integrating out $\kappa$, we 
have $C_\mu=\partial_\mu\theta$, which returns us to the original Lagrangian. 
If, however, we choose to integrate out $C$, we find instead
\be
C_\mu=g^2\epsilon_{\mu\nu}(\partial^\nu\kappa+A^\nu)
\nn\ee
and, in terms of the dual field $\kappa$, the Lagrangian becomes
\be
{\cal L}_{dual}=\frac{g^2}{2}\left(\partial \kappa + A\right)^2
\nn\ee
Notice that the presence of the $\theta F$ coupling causes 
the dual field $\kappa$ to transform transitively under the 
gauge group action,
\be
U(1):\ \ \ q\rightarrow e^{i\gamma}q\ \ \ ,\ \ \ 
\tilde{q}\rightarrow 
e^{-i\gamma}\tilde{q}\ \ \ ,\ \ \ \kappa\rightarrow\kappa+\gamma
\label{lightscamera}\ee
From a geometrical perspective, it is this transitive action which 
results 
in a stabilization of the ${\bf S}^1$ fiber at infinity, leading 
to an ALF, as opposed to ALE, space \cite{gr}. (This is discussed 
further in Appendix A. From the gauge theoretical perspective this 
construction was first described in \cite{ks}). Let us see explicitly 
how this occurs. After restricting to the vacuum moduli 
space \eqn{vacuum}, and rewriting the hypermultiplet kinetic terms 
as \eqn{kin}, the low-energy bosonic action in 
terms of the dual variable becomes,
\be
{\cal L}_{TN}=\frac{1}{2} H(r)\partial{\bf r}\cdot\partial{\bf r} 
+\frac{r}{4}(\partial\alpha+2A+\bomega\cdot\partial{\bf r})
+\frac{g^2}{2}(\partial\kappa+A)^2
\nn\ee
Recall that $\alpha=-2\arg(iq)$, and thus has period $4\pi$. It 
remains to divide by the gauge 
action \eqn{lightscamera}. There are a number of equivalent ways to 
implement this. Following \cite{gr}, we chose to set $A=0$, and to work 
in terms of the gauge invariant quantity $\psi=\alpha-2\kappa$, which 
also has period $4\pi$. We can then re-write,
\be
\frac{r}{4}(\partial\alpha+\bomega\cdot\partial{\bf r})^2
+\frac{g^2}{2}(\partial\kappa)^2&=&
\left(r+\ft12 g^2\right)\left(\partial\kappa+\ft12 r
(r+\ft12 g^2)^{-1}(\partial\psi+\bomega\cdot\partial{\bf r})\right)^2
\nn\\ && +
\frac{1}{8}H^{-1}(r)
\left(\partial\psi+\bomega\cdot\partial{\bf r}\right)^2 
\nn\ee
In this form, the second term is gauge invariant while the 
first, being a total square, is simply lost upon taking the 
$U(1)$ quotient. Thus, our final low-energy bosonic 
effective action is
\be
{\cal L}_{TN}=\ft12 H(r)\partial{\bf r}\cdot\partial{\bf r}
+\ft18 H^{-1}(r)\left(\partial\psi+\bomega\cdot\partial{\bf r}\right)^2
\nn\ee
which is indeed the sigma-model with Taub-NUT metric \eqn{tn} as advertised. 

While the previous discussion of T-duality involved only the bosonic 
fields, the extension to the supersymmetric theory is simple. A 
superfield derivation using Rocek-Verlinde transformations \cite{rv} 
in the ${\cal N}=(4,4)$ context may be found in \cite{kinky}.

\section{Worldsheet Instantons}

In the previous section we examined the gauge theory at the classical 
level and found dual descriptions of the low-energy physics. In 
one set of variables, this was the sigma-model on Taub-NUT space while, 
in the other, we found the NS5-brane, smeared on a 
transverse circle. In this section we examine the quantum corrections 
to the latter description. Before proceeding, we should make the 
usual disclaimer regarding the Coleman-Mermin-Wagner theorem 
\cite{clifford}. 
In two dimensional theories, a quantum moduli space of vacua does 
not exist, and the groundstate wavefunction spreads over all 
classical vacuum states. Here we work in the Born-Oppenheimer spirit, 
in which high momentum modes are integrated out to reveal a 
low-energy description in terms of a sigma-model on the quantum 
corrected moduli space. This approach is relevant for comparison 
to supergravity. Motivated by this comparison and, in particular, the 
expression \eqn{exp} describing a localized NS5-brane, we 
search for instanton solutions of the gauge theory. In the following 
section, we identify the relevant semi-classical 
configurations. In Section 3.2 we compute the instanton measure and 
various other accessories necessary to perform the calculation. Finally, 
in Section 3.3 we calculate the $k$-instanton contribution to the 
four-fermi vertex and translate this into the corrections to the metric.

\subsection{Aspects of Instantons}

In the gauge theory approach to sigma models, worldsheet instantons appear 
as vortices \cite{witten,schroers}. This has several advantages, among 
them the fact that vortices exist --- and contribute to correlation 
functions ---  even when there are no two-cycles in the target space 
\cite{witten,mp}. Indeed, this is the case here. Although the second 
homology class of both Taub-NUT and the smeared NS5-brane metric is 
trivial, the gauge theory does have instanton sectors labelled by,
\be
-\frac{1}{2\pi}\int F_{12} = k \in {\bf Z}
\nn\ee
where $F_{12}$ denotes the Wick rotation of the field strength $F_{01}$ 
into Euclidean space. However, although the requisite topology of the 
gauge theory exists, there are no finite action solutions to the 
equations of 
motion with the desired boundary conditions. To see this, let us 
start by choosing a classical vacuum from within the moduli 
space of vacua. We necessarily have $\phi=\sigma=0$ and, 
without loss of generality, we may employ the $SU(2)_R$ symmetry to 
further set $\tilde{q}=r^1+ir^2=0$, while
\be
|q|^2=r^3=\zeta
\label{vacuous}\ee 
The field $\theta$ is left unconstrained. To perform a semi-classical 
calculation, one must search for an instanton solution which asymptotes 
to this vacuum. However, the gauge field strength 
$F_{12}$ is a source for the massless field whose vacuum expectation 
value parameterizes the moduli space. Since this field is massless, far 
from a suitably localized 
instanton it must solve the Laplace equations in two-dimensions 
and therefore runs logarithmically at large distances. For 
this reason solutions to the equations of motion will not asymptote to 
the vacuum \eqn{vacuous} for finite $\zeta$. This problem has also 
been discussed in related models \cite{penin}. Of course, this 
behavior is unsurprising: it is simply a reflection of the 
Coleman-Mermin-Wagner theorem mentioned previously. Nevertheless, in order 
to proceed with our Born-Oppenheimer approximation, we must circumvent 
this obstacle.

Although the physics is somewhat different, formally the problem is 
reminiscent of four-dimensional instantons in Yang-Mills-Higgs 
theories. Recall that the presence of a vacuum expectation value, $v$, 
for the Higgs field implies that the Yang-Mills instanton will 
shrink to zero 
size in order to minimize its action. In this case a procedure known as 
``constrained instantons'' \cite{affleck} is employed which, in practice, 
involves expanding around the configurations which are solutions at $v=0$. 
In supersymmetric theories, these approximate solutions retain all their 
BPS properties, which allows them to deliver exact, quantitative 
information (see, for example, \cite{dkmadhm} for various applications 
and reviews). 
In the present two-dimensional situation, 
the logarithmic divergence implies that any putative solution wishes 
to expand. To halt this, we need to find the analog of the quantity 
$v$, which we can tune to zero in order to recover solutions. We will 
now show that this quantity is the asymptotic radius of Taub-NUT space, 
$g^2$. To see this, let us firstly truncate the theory to allow only 
the gauge field, $q$ and $r^3$ to vary over space-time. All 
other scalar fields are restricted to their classical expectation values 
and it can be checked that they will not further destabilize our 
solution. The equation of motion for $r^3$ in this background is, 
\be
\partial^2 r^3 = g^2e^2(|q|^2-r^3)
\label{reom}\ee
In the limit $g^2=0$, it is consistent to keep $r^3=\zeta$, even as 
$|q|^2$ moves out of vacuum. With this truncation, the bosonic Euclidean 
action is given by,
\be
S=\frac{1}{2\pi}
\int d^2x\ \frac{1}{2e^2}F_{12}^2+|{\cal D}q|^2+\frac{e^2}{2}\left(|q|^2-
\zeta\right)^2+i\theta F_{12}
\nn\ee
which is simply the abelian Higgs model at critical coupling, together with 
a $\theta$ term for the gauge field. Note the factor of $i$ which appears 
in front of this latter term after Wick rotation to Euclidean space. As is 
well known, this 
theory admits BPS vortex solutions \cite{no}. 
The first order equations of motion 
may be found by a judicious completion of the square,
\be
S =\frac{1}{2\pi}
\int d^2x\ \frac{1}{2e^2}\left(F_{12}\mp e^2(|q|^2-\zeta)\right)^2 
+|{\cal D}_1q\mp i{\cal D}_2q|^2 +(\mp\zeta+i\theta)F_{12}
\nn\ee
where the upper (lower) sign is taken for $k>0$ ($k<0$). Throughout the 
remainder of this 
paper, we work with $k>0$. Introducing the complex basis $z=x^1+ix^2$, 
our Bogomoln'yi equations take the form,
\be
F_{12}=e^2(|q|^2-\zeta)^2\ \ \ ,\ \ \ {\cal D}_zq=0
\label{bog}\ee
and solutions to these equations have the action
\be
S_k= k \zeta + k i \theta
\label{baction}\ee
%We choose to perform our semi-classical expansion about these configurations, 
%which are solutions in the $g^2\rightarrow 0$ limit. 
%To further justify the 
%importance of these configurations, we note that similar problems were 
%encountered in \cite{hk}. In this case, with ${\cal N}=(2,2)$ supersymmetry, 
%the instantons contribute to the superpotential where supersymmetry and 
%holomorphy allow the introduction of new fields which ensure the 
%existence of an instanton solution, and may then subsequently be 
%decoupled. While such an approach does not seem possible in the present 
%case, it is reassuring that, in practice, the decoupling procedure is 
%operationally equivalent to the constrained instanton approach described 
%above. 
To summarize, the instanton calculation is performed in the limit 
$g^2\rightarrow 0$, and finite $e^2$. In contrast, the sigma-model 
limit is finite $g^2$, and $e^2\rightarrow\infty$. We shall show that 
the final answer is independent of $e^2$, justfying the latter 
choice. In Appendix B, we show that the supergravity 
metric \eqn{h} yields a finite contribution to the four-fermi vertex 
of interest in the limit $g^2\rightarrow 0$,  
thus also justifying the former choice.

\subsection{The Instanton Measure}

To compute the contributions of instantons to the low-energy effective 
action, we must first isolate the zero modes. The bosonic zero modes 
are given by the solutions 
to the linearized Bogomoln'yi equations \eqn{bog},
\be
\epsilon_{\mu\nu}\partial_\mu\delta A_\nu &=& e^2(q^\dagger\delta q +q\delta 
q^\dagger) 
\nn\\
{\cal D}_\mu\delta q-i\delta A_\mu q &=& i\epsilon_{\mu\nu}{\cal D}_\nu \delta q 
+\epsilon_{\mu\nu}\delta A_\nu q
\nn\ee
which are augmented by a suitable gauge fixing condition which we take 
to be,
\be
\partial_\mu\delta A_\mu=ie^2(q\delta q^\dagger -q^\dagger \delta q)
\nn\ee
Using complex spacetime coordinates and defining $\partial=\partial_z$ and 
$\bar{\partial}=\partial_{\bar{z}}$, these can be elegantly combined into a 
bosonic Dirac equation,
\be
\Delta\left(\begin{array}{c}\delta A_z \\ \delta q\end{array}\right)=0\ \ \ 
{\rm with}\ \ \ 
\Delta=\left(\begin{array}{cc} \ft{2i}{e^2}\bar{\partial} & -q^\dagger \\ 
q & i{\cal D} \end{array}\right)
\label{delta}\ee
These equations were analyzed by E. Weinberg \cite{erick}, who showed, using 
index theory, that there 
exist $2k$ normalizable, linearly independent zero modes 
$(\delta_aA_\mu,\delta_a q)$, for 
$a=1,\ldots,2k$. Of these, two 
are Goldstone modes, arising from broken translational invariance, and given by
\be
\delta_\nu A_\mu=F_{\nu\mu}\ \ \ \ ,\ \ \ \ \delta_\nu q={\cal D}_\nu q\ \ 
\ \ \ \nu=1,2
\label{comzero}\ee
while the remaining $2(k-1)$ are not generated by any symmetry. 
It can be shown that these remaining zero modes correspond to the 
decomposition of the $k$-vortex soliton into $k$ single vortices
with arbitrary positions given by the zeroes of the Higgs field 
\cite{taubes}. We can introduce $2k$ collective 
coordinates, $X^a$, $a=1,\ldots,2k$, which parameterize the multi-vortex 
moduli space ${\cal M}_k$. 
This space is endowed with a complete K\"ahler metric, given by the overlap of 
the zero modes \cite{samols}, 
\be
g_{ab}=\frac{1}{2\pi}
\int d^2x\ \frac{1}{2e^2}\delta_aA_\mu\,\delta_bA_\mu+\frac{1}{2}\delta_a q\,
\delta_b q^\dagger 
+\frac{1}{2}\delta_a q^\dagger\,\delta_b q
\label{vortmet}\ee
where the complex structure of $g$ is inherited from the complex structure 
of the two-dimensional 
worldsheet.
%with the complex structure $J$ given by,
%\be
%J_a^{\ b}\delta_bA_\mu=\epsilon_{\mu\nu}\delta_aA_\nu\ \ \ \ ,\ \ \ \ 
%J_a^{\ b}\delta_bq=i\delta_aq
%\nn\ee
The moduli space ${\cal M}_k$ decomposes metrically as 
\be
{\cal M}_k={\bf R}^2\times\tilde{\cal M}_k
\nn\ee
where ${\bf R}^2$ is parameterized by $X^\mu$, the center of mass of the 
vortices. The 
restriction of the metric $g$ to the ${\bf R}^2$ factor can be 
easily calculated by 
substituting the zero modes \eqn{comzero} into the metric \eqn{vortmet}. 
It yields $\zeta k\delta_{\mu\nu}$, as expected for a soliton of 
``mass'' $\zeta k$ (``mass'' becomes ``action'' for instantons). 
The centered moduli space 
$\tilde{\cal M}_k$ is parameterized by the relative positions $Y^p$, 
$p=1\ldots,2(k-1)$ of $k$ vortices.  
We denote the metric on this $2(k-1)$ dimensional space as 
$\tilde{g}_{pq}$. Its 
analytic form is unknown, even for $k=2$. 

When performing the instanton 
calculation we must integrate over all these collective coordinates. 
The measure is obtained by changing variables in the path integral, 
and is given by \cite{bernard},
\be
\int d\mu_B=\frac{\zeta k}{2\pi} \int d^2X\,\prod_{p=1}^{2(k-1)}dY^p\ 
\frac{\det^{1/2}(\tilde{g})}{(2\pi)^{k-1}}
\label{dmub}\ee

\subsubsection*{\rm {\em Fermion Zero Modes}}

The fermionic zero modes are related to the bosonic zero modes via 
the unbroken supersymmetry. To see this more explicitly, we examine 
the Dirac equations for the vector and hypermultiplet fermions in 
vortex background, in the limit $g^2\rightarrow 0$,
\be
\Delta\left(\begin{array}{c} i\bar{\lambda}_+/\sqrt{2} \\ \psi_- 
\end{array}\right)&=&\Delta\left(\begin{array}{c} 
\tilde{\lambda}_+/\sqrt{2} \\ \bar{\tilde{\psi}}_- \end{array}\right) =0 
\nn\\
\Delta^\dagger\left(\begin{array}{c} i\bar{\lambda}_-/\sqrt{2} \\ \psi_+
\end{array}\right)&=&\Delta^\dagger\left(\begin{array}{c} - 
\tilde{\lambda}_-/\sqrt{2} \\ \bar{\tilde{\psi}}_+\end{array}\right)=0
\nn\ee
where the Dirac operator $\Delta$ is the same as that encountered in the 
analysis of the bosonic zero modes \eqn{delta}. The fermionic 
zero modes are related to the bosonic ones by 
$\lambda\sim\sqrt{2}\delta A_z$ and $\psi\sim\delta q$. 
Note that, while $\Delta$ has $2k$ zero modes, $\Delta^\dagger$ has none. 
To see this consider the action on an arbitrary complex doublet $Y$ and 
define the norm,
\be
\|\Delta^\dagger Y \| &=&
|\ft{2i}{e^2}\partial y_1+q^\dagger y_2|^2+\ft{2}{e^2}|
i\bar{\cal D}y_2-qy_1|^2 \nn\\
&=& |\ft{2}{e^2}\partial y_1|^2+\ft{2}{e^2}|\bar{\cal D}y_2|^2+\ft{2}{e^2}|qy_1|^2
+|q^\dagger y_2|^2-\ft{2i}{e^2}\left(y_1y_2^\dagger{\cal D}q-y_1^\dagger y_2\bar{\cal D}
q^\dagger\right)
\nn\ee
The last two terms vanish when evaluated on the background of the vortex, 
while the middle two ensure that $\Delta^\dagger Y=0$ if and only if 
$Y=0$. Thus the zero modes 
are carried by the pairs $(i\bar{\lambda}_+,\psi_-)^T$, 
$(\tilde{\lambda}_+,\bar{\tilde{\psi}}_-)^T$, $(i\lambda_-,\bar{\psi}_+)^T$, 
and $(-\bar{\tilde{\lambda}}_-,\tilde{\psi}_+)^T$. For example, the 
Goldstone bosons of equation \eqn{comzero} are related to the 
fermionic zero modes generated by the four broken supersymmetries,
\be
\begin{array}{cccc}
\bar{\lambda}_+=\ft{1}{\sqrt{2}}F_{12}\alpha_1, & 
\lambda_-=\ft{1}{\sqrt{2}}F_{12}\alpha_2, & 
\tilde{\lambda}_+=\ft{1}{\sqrt{2}}F_{12}\tilde{\alpha}_1, &  
\bar{\tilde{\lambda}}_-=\ft{1}{\sqrt{2}}F_{12}\tilde{\alpha}_2 \\
\psi_-=\bar{\cal D}q\alpha_1, & \bar{\psi}_+=\bar{\cal D}q\alpha_2, 
& \bar{\tilde{\psi}}_-=\bar{\cal D}q\tilde{\alpha}_1 , &  
\tilde{\psi}_+=\bar{\cal D}q\tilde{\alpha}_2\end{array}
\label{fzero}\ee
The fermionic measure for these broken supersymmetries is determined 
by calculating the overlap as for the bosonic case\footnote{The constant 
$\ft12\zeta k$ differs by a factor of $1/2$ from normalization of the 
bosonic Goldstone modes. This can be traced to the fact that, under 
supersymmetry transformation, $\delta\psi=\sqrt{2}i\bar{\cal D}q\xi$, 
and thus $\alpha=\sqrt{2}\xi$, where $\xi$ are the infinitesimal 
supersymmetry parameters}, yielding
\be
\int d\bar{\mu}_F=\int d^2\alpha\, d^2\tilde{\alpha}\ (\ft12\zeta k)^{-2}
\label{dmufbar}\ee
There are $4(k-1)$ further fermionic zero modes, related by unbroken 
supersymmetry to the $2(k-1)$ relative vortex positions. Let us 
denote the corresponding Grassmannian collective coordinates as 
$\beta^p$ and $\tilde{\beta}^p$, with $p=1,\ldots,2(k-1)$, where 
the $\beta$'s arise from the $(\lambda,\psi)$ pairs, and the 
$\tilde{\beta}$'s from the $(\tilde{\lambda},\tilde{\psi})$ pairs. As 
with their bosonic partners, each of these must also be 
integrated over, with the measure given by \cite{dkm}
\be
\int d\tilde{\mu}_F =\int\prod_{p=1}^{2(k-1)}\,d\beta^p d\tilde{\beta}^p
\frac{1}{\det(\tilde{g})}
\label{dmuftilde}\ee

\subsubsection*{\rm{\em The Action, Determinants and Long-Distance 
Behavior}}

While the constant part of the instanton action is given by \eqn{baction}, 
the instanton action can, in principle, also depend on the collective 
coordinates. This occurs when the zero modes discussed in the last section, 
each of which solves the linearized equations of 
motion, cannot all be simultaneously integrated to solutions of the 
full equations 
of motion. While this does not occur for the bosonic collective 
coordinates \cite{taubes}, it is commonplace 
for fermionic collective coordinates and, indeed, is even required by 
supersymmetry considerations \cite{dkm}. (For a toy model, and a 
discussion of these issues, see \cite{stefan}).  To see this, note that 
the vortices discussed here preserve ${\cal N}=(2,2)$ 
non-chiral\footnote{Non-chiral supersymmetry in space-time 
dimensions less than 
two, refers to the dimensional reduction of a non-chiral theory in two 
dimensions. 
To see that the theory on the vortex worldvolume is indeed non-chiral in 
the two-dimensional sense, it suffices to notice that the gauge theory 
described in Section 2 can be lifted in a supersymmetric fashion to 5+1 
dimensions, where the vortices have a 3+1 dimensional worldvolume} 
supersymmetry on their worldvolume. The low-energy dynamics of these 
objects is described by a  ``0+0-dimensional sigma-model''. Naively, one 
may imagine that, since sigma models have only derivative couplings, such 
an action is trivial. However, the non-chiral supersymmetric extensions 
of sigma-models also include non-derivative, four-fermi 
couplings \cite{dan}. These survive in the instanton action and, in the 
present case, are given by \cite{dkm}
\be
S_{4-fermi}=\ft14 \tilde{R}_{pqrs}\beta^p\beta^q\tilde{\beta}^r
\tilde{\beta}^s
\label{fourfermi}\ee
where $\tilde{R}$ is the Riemann tensor on the relative vortex 
moduli space $\tilde{\cal M}_{k}$.

In any instanton calculation, one must also integrate over 
the non-zero modes. In supersymmetric theories, the non-zero eigenvalues 
of the bosonic and fermionic operators around the background of a BPS 
instanton are guaranteed to coincide. In the case of four dimensional 
instantons, 't Hooft showed long ago that this is sufficient to 
ensure the cancellation of the one-loop determinants. However, this 
cancellation need not necessarily occur for operators with continuous 
spectra for, while the spectrum of eigenvalues must coincide, the 
density need not. Indeed, for instantons in three-dimensional gauge 
theories, it can be shown that the integration over non-zero modes 
leads to a finite, calculable contribution \cite{early}. In the present 
case however, such fears are groundless as the exponential fall-off 
of the vortex tail ensures that the spectrum is suitably well-behaved  
\cite{erick} and the non-zero modes cancel. 

Finally, we turn to the long-distance behavior of fields in the 
background of the instanton. While no explicit analytic solutions 
to the Bogomoln'yi equations \eqn{bog} have been found, at distances 
large compared to all other length scales, the asymptotic form of 
the solutions is known \cite{vega}. Using polar coordinates, 
$z=\rho\exp(i\vartheta)$, the solution for a $k$-vortex configuration 
with center of mass at the origin, $X=0$, becomes
%solution \cite{taubes}. Using polar coordinates, 
%$z=\rho\exp(i\psi)$, we have the ansatz,
%\be
%q=\sqrt{\zeta}e^{ik\psi}f(\rho)\ \ \ ,\ \ \ A_\rho=0\ \ \ ,\ \ \ 
%A_\psi=k a(\rho)
%\nn\ee
%Substituting this ansatz into the Bogomoln'yi equations \eqn{bog}, we arrive 
%at the pair of well-known coupled first order differential equations,
%\be
%\rho \frac{df}{d\rho}+k(1-a)f=0\ \ \ ,\ \ \ \frac{k}{\rho}\frac{da}{d\rho} 
%=e^2\zeta(f^2-1)
%\nn\ee
%These equations were analyzed in detail in \cite{vega} and, although no 
%exact solutions were found, the asymptotic form of the solution is known 
%to be,
\be
|q|^2\rightarrow\zeta\left(1-l_k(Y^p,\vartheta)
\sqrt{\frac{2\pi L}{\rho}}\exp(-\rho/L)\right)
\label{qlarge}\ee
where the characteristic length scale of the vortex is 
\be
L=\frac{1}{\sqrt{2e^2\zeta}}
\nn\ee
The functions $l_k(Y^p,\vartheta)$ characterize the exponential 
return to vacuum. The fact that the vortex tail is exponential, 
as opposed to polynomial, implies that these coefficients 
are functions of both the relative positions of the 
vortices, as well as the angular position on the complex plane, 
as indicated. Certain properties of these functions were recently 
studied in \cite{nickmartin}.  
For a single vortex, $l_1$ is simply a numerical coefficient. It is not 
known analytically, but has been calculated numerically 
\cite{vega,speight}. The newer result of Speight \cite{speight} 
is\footnote{To compare 
with the conventions of \cite{speight}, one must multiply by 
$2\pi$. The older numerical result of \cite{vega} gives 
$l_1\approx 1.708$.}
\be
l_1\approx 1.683\pm 0.001
\label{nastylittlenumber}\ee
%For the rotationally symmetric $k$-vortex solution, in which the 
%Higgs field has only a single zero of degeneracy $k$ and $Y^p=0$, 
%a rather peculiar bound on their behavior can be derived \cite{vega}
%\be
%l_k(Y^p=0)>\frac{3}{2}k+\left(\frac{\pi}{3\sqrt{3}}-\frac{1}{2}\right)k^2
%\nn\ee

\subsection{The Calculation}

We will now collect together all the pieces in order to compute the 
$k$-instanton contribution to the four-fermi correlation 
function,
\be
G_4^{(k)}(x_1,x_2,x_3,x_4)&=&\left.\langle\bar{\psi}_+(x_1)\psi_-(x_2)
\tilde{\psi}_+(x_3)\bar{\tilde{\psi}}_-(x_4)\rangle\right|_{k-instanton} \nn\\
&=&\int d\mu_B d\bar{\mu}_F d\tilde{\mu}_F
\ \bar{\psi}_+(x_1)\psi_-(x_2)
\tilde{\psi}_+(x_3)\bar{\tilde{\psi}}_-(x_4)\ e^{-S_k-S_{4-fermi}}
\nn\ee
where the various components of this expression can be 
found in equations \eqn{baction}, \eqn{dmub}, \eqn{dmufbar}, 
\eqn{dmuftilde} and \eqn{fourfermi}. The fermions are replaced in the 
path-integral by their zero-mode values in the $k$-instanton background. 
In order to determine instanton contributions to higher dimension 
operators, one would need to know the explicit form for all 
such zero modes. However, for the special case of the 
four-fermi correlation function $G_4$, we need only the expression for 
the zero modes arising from unbroken supersymmetry \eqn{fzero}. This 
is the semi-classical reflection of the non-renormalization 
theorems which ensure the calculability of two-derivative terms 
in theories with eight supercharges. From equations \eqn{fzero} 
and \eqn{qlarge}, we have the large distance expansion
\be
\psi(x_i)&\rightarrow& l_k(Y^p,\vartheta-\vartheta_i)
\sqrt{\frac{\pi\zeta}{2L\rho}}
e^{-\rho/L}e^{-i(k-1)(\vartheta-\vartheta_i)}\alpha 
\nn\\ 
&=& l_k(Y^p,\vartheta-\vartheta_i)
\sqrt{\zeta}e^{-i(k-1)(\vartheta-\vartheta_i)}S_F \alpha
\nn\ee
Here $S_F$ is the leading order behavior of the diagonal component of 
the fermionic propagator for a Dirac fermion of mass $1/L$ in the vacuum, 
\be
S_F\rightarrow \sqrt{\frac{\pi}{2L\rho}}\exp(-\rho/L)
\nn\ee
as befits a Green's function for the Dirac operator $\Delta/2\pi$, 
defined in \eqn{delta}. 
Note that all dependence on the gauge coupling constant $e^2$ 
appears through the vortex length scale $L$ which, in turn, 
appears only in the propagator. This is to be expected since 
gauge coupling constants cannot appear in the metric of the 
Higgs branch \cite{aps}. Note, in contrast, that not all position 
dependence can be absorbed in the propogator; angular dependence 
remains. This results in a form-factor for $G_4^{(k)}$ with $k>1$ 
which can be converted to higher derivative interactions. To leading 
order in the derivative expansion, we have
\be
G_4^{(k)}(x_1,x_2,x_3,x_4) = 2\zeta\,k^{-1}\pi^{-1}\, 
\nu(\tilde{M}_k)\,e^{-\zeta k-ik\theta}\int d^2X 
\prod_{i=1}^4S_F(X-x_i)
\label{answer}\ee
where the integration over all relative bosonic 
and fermionic zero modes has been collected into the function $\nu$, 
and expressed as an integral over the $d=2(k-1)$ dimensional 
moduli space $\tilde{\cal M}_k$ as in \cite{dkm}, 
\be
\nu(\tilde{M}_k) &=& \frac{1}{(2\pi)^{d/2}} 
\int \prod_{p=1}^{d} dY^pd\beta^pd \tilde{\beta}^p d\vartheta\,   
\frac{l_k^4(Y^p,\vartheta)}{\sqrt{\det(\tilde{g})}}  
\frac{e^{-4i(k-1)\vartheta}}{2\pi}
\exp\left(-\frac{1}{4} \tilde{R}_{pqrs}\beta^p\beta^q\tilde{\beta}^r
\tilde{\beta}^s\right)\nn\\
&=& \frac{1}{(8\pi)^{d/2}(d/2)!}\int
\frac{\prod_{p=1}^{d}dY^p}{\sqrt{\det(\tilde{g})}}\,
\epsilon^{p_1p_2\ldots p_{d}}
\epsilon^{q_1q_2\ldots q_{d}}
\ \tilde{R}_{p_1p_2q_1q_2}\ldots 
\tilde{R}_{p_{d-1}p_{d}q_{d-1}q_{d}}\nn\\
&&\ \ \ \ \ \ \ \ \ \  \ \ \ \ \ \ \ \ \times\frac{1}{2\pi}\int d\vartheta
\ l_k^4(Y^p,\vartheta)\,e^{-4i(k-1)\vartheta}
\label{arnold}\ee
For $k=1$, we have $\nu_1=l_1^4$. For $k>1$, the expression above 
is familiar as the integral of the Euler form over the $k$-vortex 
relative moduli space, $\tilde{\cal M}_k$, weighted by a specific 
Fourier mode of $l_k^4(Y^p,\vartheta)$, the function which 
characterizes the exponential fall-off 
of the vortex solution.

Summing over all instanton sectors, $k\in{\bf Z}$, the correlation 
function   
$\sum_kG_4^{(k)}$ is equivalent to a four-fermi vertex for $\psi$ in the 
low-energy effective action. Imposing the constraints \eqn{chipsi} 
on this low-energy coupling, and re-expressing $\zeta\equiv r$, 
allows us to re-write the interaction in terms of the massless 
fermions $\chi$, 
\be
\sum_{k=1}^\infty \xi_k
\left(e^{-ik\theta} 
\bar{\tilde{\chi}}_+\tilde{\chi}_-\bar{\chi}_+\chi_-
+e^{+ik\theta}{\tilde{\chi}}_+\bar{\tilde{\chi}}_-{\chi}_+
\bar{\chi}_-\right)
\label{okey}\ee
where the coefficient is given by
\be
\xi_k = \frac{1}{2^5 \pi r k}\  
\nu(\tilde{\cal M}_k)\ e^{-kr}
\nn\ee
To compare to the supergravity prediction \eqn{exp}, we 
examine the low-energy action \eqn{compare}. The coefficient in front 
of the four-fermi term above is related to the Riemann tensor 
of the target space metric, computed with respect to the 
connection with torsion. In Appendix B we compute this 
object, at leading order in $1/r$, and to leading order 
in $g^2$, as is warranted given the discussion of Section 3.1. 
Taking into account the symmetries of the Riemann tensor, the prediction 
for the low-energy four-fermi term has the same functional form as 
\eqn{okey}, but with the coefficient 
$\xi_k$ replaced by $\tilde{\xi}_k$, given by
\be
\tilde{\xi}_k=\frac{k^2}{4\pi r}\ e^{-kr}
\nn\ee
We thus see that agreement with supergravity requires 
$\nu(\tilde{M}_k)=8k^3$. 
It would be interesting to recover this prediction solely 
from the study of vortices. For now we claim agreement only for the 
$k=1$ instanton sector, where we have $\nu_1=l_1^4$. The 
supergravity prediction requires
\be
l_1=8^{1/4}\approx 1.682
\nn\ee
Encouragingly, this is in accord with the value \eqn{nastylittlenumber} 
obtained through numerical studies of the vortex equations 
\cite{vega,speight}. 

\section*{Acknowledgements}

I'd like to express my thanks to Luis Bettencourt, Neil Constable, 
Nick Dorey, Ansar Fayyazuddin, Dan Freedman, Ami Hanany, 
Conor Houghton, Clifford Johnson, Neil Lambert, Nick Manton, 
Adam Ritz, Martin Speight, Brett Taylor, Wati Taylor, Jan Troost and 
Piljin Yi for useful discussions, and especially to 
Mina Aganagic, Kentaro Hori and Andreas Karch for collaboration 
in the initial stages of this 
project. My thanks also to Clifford Johnson and Andreas Karch for 
correcting an important mistake in the original version of these 
acknowledgements. 
I'm supported by a Pappalardo fellowship, and am very grateful 
to Neil Pappalardo for the money. This work was also supported in part 
by funds provided by the U.S. Department of Energy (D.O.E.) under 
cooperative research agreement \#DF-FC02-94ER40818.

\appendix{Linear Sigma-Model for ALF Spaces}

While the gauged linear sigma-model model for asymptotically 
locally Euclidean (ALE) spaces is well known, the deformation 
to asymptotically locally flat (ALF) spaces appears to be less 
familiar to string theorists. For this reason we include this appendix 
describing the gauge theory whose Higgs branch is 
endowed with the hyperk\"ahler metric on the 
ALF space with $A_{N-1}$ singularity, given by,
\be
ds^2=H(r)d{\bf r}\cdot d{\bf r}+\ft14 H(r)^{-1}\left(d\psi + \bomega
\cdot d{\bf r}\right)^2
\label{ohdear}\ee
where ${\bf r}$ is a three-vector, and $\psi$ has period $4\pi$. The 
connection $\vec{\omega}$ is determined by 
$\nabla\times\vec{\omega}=2\nabla H$, where $H$ is the harmonic function,
\be
H(r) = \frac{1}{g^2} + \frac{N}{2r}
\label{alf}\ee
For $N=1$, this is the Taub-NUT metric of equation \eqn{tn}. The metric 
is written in Gibbons-Hawking coordinates which describe the ALF space 
as a fibration of ${\bf S}^1$ --- parameterized by $\psi$ --- over 
an ${\bf R}^3$ base. Asymptotically ${\bf S}^1$ has radius $g$, 
and provides a Hopf fibration over ${\bf S}^2=\partial{\bf R}^3$, with 
winding number $N$. Rotations around this ${\bf S}^1$ yield a 
tri-holomorphic isometry which we denote as $U(1)_F$,
\be
U(1)_F:\quad\quad \psi\rightarrow\psi+\alpha
\label{steve}\ee
It is instructive to consider the limit $g^2\rightarrow\infty$, in 
which the boundary becomes the Lens space 
${\bf S}^3/{\bf Z}_N$, and \eqn{alf} is simply the flat metric on the 
ALE orbifold ${\bf C}^2/{\bf Z}_N$. We will refer to this as the 
ALE limit. In this case, there is a well-known gauged linear sigma 
model which reproduces this target space. It appears naturally in 
string theory 
as the theory on a probe D-brane \cite{dm}. It is the quiver theory 
with 8 supercharges 
--- corresponding to ${\cal N}=(4,4)$ supersymmetry in two dimensions 
---  associated to the affine $A_{N-1}$ Dynkin diagram. The theory has 
gauge group ${\cal G}=\prod_{i=1}^N U(k_i)$ where, 
for a single D-brane, $k_i=1$ for all $i$.  The matter 
content consists of a bi-fundamental hypermultiplet transforming under 
each pair of adjacent gauge groups, $(+k_i,-k_{i+1})$, where the index 
$i$ is defined modulo $N$. Since the overall ``center-of-mass'' $U(1)$ 
decouples, the interacting matter is,
\begin{center}
$A_{N-1}$ ALE Theory:\ \ $U(1)^{N-1}$ with $N$ hypermultiplets
\end{center}
The Higgs branch of this theory reproduces the ${\bf C}^2/{\bf Z}_N$ 
orbifold. The bridge between the gauge theory and the geometry is 
provided by the hyperkahler quotient construction acting on the 
quaternionic space ${\bf H}^N$, parameterised by the $N$ complex doublets 
$w$, which are the scalar components of the hypermultiplets. In 
the notation of Section 2, we have $\omega_i^\dagger = 
(q_i^\dagger,\tilde{q}_i)$. The moment 
maps coincide with the D-terms of the gauge theory and are given by,
\be
\bmu_i=\omega^{\dagger}_i\btau\omega_i
-\omega_{i+1}^\dagger\btau\omega_{i+1} 
\label{charlotte}\ee
where $\btau$ are the Pauli matrices. 
Note that the sum of the moment maps is trivial, leaving only $(N-1)$ 
linearly independent triplets of constraints. After dividing by the 
$U(1)^{N-1}$ gauge action,  we arrive at the 
metric \eqn{alf} in the ALE limit, $1/g^2=0$. It is important to 
note that the action of the $U(1)_F$ isometry on the target space 
\eqn{steve} arises from the flavor 
symmetry of the gauge theory,
\be
U(1)_F:\quad\quad\omega_i\rightarrow\exp(i\alpha)\,\omega_i\ \ \ \ \ \ 
\forall \ i
\label{sophie}\ee
We would like to generalise this gauged linear sigma model to the 
ALF metric, with finite $g^2$. Geometrically, this requires squashing 
the ${\bf S}^1$ fibre at infinity. The hyperk\"ahler 
quotient construction for such a space was discussed in \cite{gr}, 
and a gauge theoretic interpretation (in the three-dimensional 
context) was given in \cite{ks}. The upshot of these papers is that 
the squashing from the ALE to ALF space can be achieved in a 
two-step process. Firstly, one gauges the $U(1)_F$ isometry; secondly, 
this is coupled to a ``linear multiplet''. In the $d=1+1$ dimensions 
of interest, a linear multiplet is also referred to as a twisted 
hypermultiplet\footnote{In $d=2+1$ dimensions, the linear multiplet 
is also known as a twisted vector multiplet, and couples to $U(1)_F$ 
through a Chern-Simons interaction. In $d=3+1$, the linear multiplet 
contains a two-form field and three scalars, and couples to $U(1)_F$ 
through a ``BF'' interaction.}. Thus the matter content for the 
gauged linear sigma-model describing the ALF space is,
\begin{center}
$A_{N-1}$ ALF Theory:\ \ $U(1)^{N}$ with $N$ hypermultiplets and 1 
twisted hypermultiplet
\end{center}
For $N=1$ this is the theory described in Section 2. Let us denote the 
gauge field appearing in the additional vector multiplet as $A_F$. 
We further decompose the four scalars of the twisted hypermultiplet 
as a triplet ${\bf r}$ and a singlet $\theta$. The full couplings 
between the vector and twisted hypermultiplet is given in Section 2.1. 
Here we isolate the terms relevant for the hyperk\"ahler quotient. 
They include the terms,
\be
\Delta{\cal L}=\frac{1}{2e^2} dA_F^2 +
\frac{1}{2g^2}\left(d\theta^2+d{\bf r}^2\right)+{\theta}\wedge dA_F
\label{hugo}\ee 
We have introduced the coupling constants $e^2$ and $g^2$. The metric 
on the Higgs branch cannot depend upon coupling constants for vector 
multiplets \cite{aps} and, to truly restrict to the Higgs branch, we 
take the $e^2\rightarrow\infty$ limit. However, the metric on the 
Higgs branch does depend on $g^2$. Indeed, this will determine the 
amount of asymptotic squashing of the ${\bf S}^1$ fiber through its 
contribution to the harmonic function \eqn{alf}. The key to seeing this 
fact lies in understanding the supersymmetric completion of 
$\theta\wedge dA_F$ term in \eqn{hugo}. As well as various fermion 
couplings, there are also further D-terms which enhance the moment map 
for the $U(1)_F$ gauge action to include a coupling to ${\bf r}$, 
\be
\bmu_F=\sum_{i=1}^N w^\dagger_i\btau w_i-{\bf r}
\label{arthur}\ee 
The presence of the ${\bf r}$ term in the moment map implies the presence 
of a field transforming transitively under $U(1)_F$ \cite{gr}. As explained 
in detail in Section 2, such a field exists, but appears only after dualizing 
$\theta$ in exchange for a new scalar field $\kappa$ through the relationship 
$d\theta=g^2{}^\star (d\kappa + A)$, after which the kinetic terms require, 
\be
U(1)_F:\quad\quad\sigma\rightarrow\sigma+ \alpha
\label{fred}\ee
The explicit hyperk\"ahler quotient reduction with this moment map was performed 
in \cite{gr} where it was shown to reproduce the metric \eqn{ohdear}. The case of 
$N=1$ was described in detail in Section 2.2.

\appendix{The Riemann Tensor}
%\label{app:B}

In this appendix we calculate the leading order contribution from 
all sectors of instantons and anti-instantons to the Riemann 
tensor for the metric
\be
ds^2=H(r,\theta)=H(r)\left(d\bf{r}\cdot d{\bf r} +d\theta^2\right)
\nn\ee
where ${\bf r}=(r^1,r^2,r^3)$, and $H$ depends only on $\theta$ and 
$|{\bf r}|=r$, 
\be
H(r,\theta)=\frac{1}{g^2}+\frac{1}{2r}
\frac{\sinh r}{\cosh r -\cos \theta}
\label{apph}\ee
Setting $r^4\equiv\theta$, the torsion is given by
\be
T_{ijk}=\epsilon_{ijk}^{\ \ \ l}\partial_l H^{-1}
\label{appt}\ee
We define the vierbein one-forms $e^\alpha=e^\alpha_i dr^i$,
\be
e^\alpha=H^{1/2}dr^i\delta^\alpha_i
\nn\ee
which facilitates calculation of the spin connection, $\omega$
\be
\omega^\alpha_{\ \beta}=\ft12(H^{-1}\partial_jH\, dr^i-T^i_{\ jk}dr^k)
\delta_i^\alpha\delta^j_\beta-\ft12H^{-1}\partial_iH\,dr^j\delta^{i\alpha}
\delta_{j\beta}
\nn\ee
It can be checked that these reproduce the Levi-Civita connection 
when $T\equiv 0$, and satisfy Cartan's first structure equation,
\be
de^\alpha+\omega^\alpha_{\ \beta}\wedge e^\beta = T^\alpha\equiv\ft12 
T^\alpha_{\ \beta\gamma}e^\beta\wedge e^\gamma
\nn\ee
The curvature two-forms are now determined using Cartan's second 
equation,
\be
d\omega^\alpha_{\ \beta}+\omega^\alpha_{\ \gamma}\wedge 
\omega^\gamma_{\ \beta}=R^\alpha_{\ \beta}\equiv\ft12 
R^\alpha_{\ \beta\gamma\delta}e^\gamma\wedge e^\delta
\nn\ee
From which we may extract the Riemann tensor which, in the original 
$r^i$ coordinates, reads
\be
R^i_{\ jkl}&=& \left(\ft14H^{-2}\partial_jH\partial_{[l}H-\ft14 
H^{-1}\partial_k H T^k_{\ j[l}-\ft12\partial_j(H^{-1}\partial_{[l}
H)\right)\delta^i_{k]} \nn\\
&& -\left(\ft14H^{-2}\partial_iH\partial_{[l}H+\ft14H^{-1}
\partial_kHT^i_{ m[l}\delta^{km}-\ft12\partial_i(H^{-1}\partial_{[l}H)
\right)\delta_{k]j}\nn\\ &&
-\ft12H^{-1}T^i_{\ lk}\partial_jH+\ft14T^m_{\ k[j}\delta_{l]m}
H^{-1}\partial_iH+\ft12\partial_{[l}T^i_{\ k]j}-\ft14T^i_{\ m[l}
T^m_{\ k]j} \nn\\ && -\ft12 H^{-2}(\partial_m H)^2\delta^i_{[k}\delta_{j]l}
\nn\ee
To compare with the instanton calculation, we should expand this 
expression for large $r$. In fact, from \eqn{apph}, we see that 
there exist two such expansion paramters: $r^{-1}$ and $e^{-r}$. 
We wish to keep all orders in the instanton expansion, but only 
leading order in $r^{-1}$. Expanding the function \eqn{apph}, 
we have
\be
\partial H \rightarrow \frac{1}{2r}
\sum_{k=1}^\infty\sum_{\pm} k\,
e^{-kr \pm ik\theta}
\ee
from which we deduce that the leading order contribution 
to the Riemann tensor appears at $1/r$, and is given by the 
eminently more managable
\be
R^i_{\ jkl}\rightarrow-\ft12 H^{-1}\partial_j\partial_{[l}H
\delta_{k]}^i+\ft12H^{-1}\partial_i\partial_{[l}H\delta_{k]j}
-\ft12\partial_{[l}T^i_{\ k]j}
\nn\ee
Finally, following the discussion of the constrained instantons 
presented in Section 3, we also wish to expand the Riemann tensor 
in powers of $g^2 \ll 1$. Since the first two terms are of 
order $g^2$, while the third is of order $g^6$, we neglect the 
torsion completely: 
\be
R^i_{\ jkl}\rightarrow-\ft12 H^{-1}\partial_j\partial_{[l}H
\delta_{k]}^i+\ft12H^{-1}\partial_i\partial_{[l}H\delta_{k]j}
\nn\ee
To compare with the instanton computation, we change to 
a complex basis of coordinates,
\be
z^1=r^1+ir^2\ \ \ ,\ \ \ z^2=r^3+i\theta
\nn\ee
and compute the components of the Riemann tensor evaluated at the 
point $z^1=0$. At leading order in $r^{-1}$ and $g^2$, we find
\be
\left.R^{\bar 1}_{\ 212}\right|_{z^1=0} &=& 
\left.R^{\bar{1}}_{\ 2\bar{1}\bar{2}}\right|_{z^1=0}\rightarrow 0 \nn\\
\left.R^{1}_{\ 212}\right|_{z^1=0} &\rightarrow& -\frac{g^2}{2r}
\sum_{k=1}^\infty k^2e^{-kr-ik\theta} \label{appr} \\ 
\left.R^{\bar{1}}_{\ \bar{2}\bar{1}\bar{2}}\right|_{z^1=0} 
&\rightarrow& -\frac{g^2}{2r}\sum_{k=1}^\infty k^2
e^{- kr+ ik\theta}
\nn\ee 
To compare with the appearance of this Riemann tensor in the 
low-energy effective action \eqn{compare}, we must lower the 
sole upper index. At leading order in $r$, this simply removes 
the factor of $g^2$, and the relevant components of the tensor are,
\be
\left.R_{\bar{1}212}\right|_{z^1=0}&\rightarrow& -\frac{1}{2r}
\sum_{k=1}^\infty k^2 e^{-kr-ik\theta} \nn\\ 
\left.R_{{1}\bar{2}\bar{1}\bar{2}}\right|_{z^1=0}
&\rightarrow& -\frac{1}{2r}
\sum_{k=1}^\infty k^2 e^{-kr+ik\theta}
\nn\ee


\newpage

\begin{thebibliography}{99}

\small
\parskip=0pt plus 2pt

\bibitem{ov} H. Ooguri and C. Vafa, ``{\em Two-Dimensional Black 
Hole and Singularities of CY Manifolds}'', Nucl. Phys. {\bf B463} 
(1996) 55, hep-th/9511164. 
%%CITATION = HEP-TH 9511164;%%
\bibitem{buscher} T. Buscher, ``{\em A Symmetry of the String 
Background Field Equations}'', Phys. Lett. {\bf B184} (1987) 
59; ``{\em Path-Integral Derivation of Quantum Duality in 
Non-Linear Sigma Models}'', Phys. Lett. {\bf B201} (1988) 
466.
%%CITATION = PHLTA,B194,59;%%
%%CITATION = PHLTA,B201,466;%%
\bibitem{tduality} A. Giveon, M. Porrati and E. Rabinovici, 
``{\em Target Space Duality in String Theory}'', Phys. Rept. 
{\bf 244} (1994) 77, hep-th/9401139.
%%CITATION = HEP-TH 9401139;%%
\bibitem{witten} E. Witten, ``{\em Phases of N=2 Theories in Two 
Dimensions}'', Nucl. Phys. {\bf B403} (1993) 159, hep-th/9301042. 
%%CITATION = HEP-TH 9301042;%%
\bibitem{mp} D. Morrison and M.R. Plesser, ``{\em Summing the Instantons: 
Quantum Cohomology and Mirror Symmetry in Toric Varieties}'', 
Nucl. Phys. {\bf B440} (1995) 279, hep-th/9412236.
%%CITATION = HEP-TH 9412236;%%
\bibitem{hv} K. Hori and C. Vafa, ``{\em Mirror Symmetry}'', 
hep-th/0002222.
%%CITATION = HEP-TH 0002222;%%
\bibitem{hk} K. Hori and A. Kapustin, ``{\em Duality of the 
Fermionic 2d Black Hole and N=2 Liouville Theory as Mirror Symmetry}'', 
JHEP {\bf 0108} (2001) 045, hep-th/0104202. \\
%%CITATION = HEP-TH 0104202;%%
K.~Hori and A.~Kapustin, 
``{\em Worldsheet Descriptions of Wrapped NS Five-Branes}'', 
hep-th/0203147.
%%CITATION = HEP-TH 0203147;%%
\bibitem{chs} C. Callan, J. Harvey and A. Strominger,
``{\em World Sheet Approach To Heterotic Instantons And Solitons}'', 
Nucl.\ Phys.\ {\bf B359} (1991) 611.
%%CITATION = NUPHA,B359,61
\bibitem{ghm} R. Gregory, J. Harvey and G. Moore, 
``{\em Unwinding strings and T-duality of Kaluza-Klein and 
H-Monopoles}'', Adv. Theor. Math. Phys. {\bf 1} (1997) 283, 
hep-th/9708086.
%%CITATION = HEP-TH 9708086;%%
\bibitem{affleck} I.~Affleck, ``{\em On Constrained Instantons}'', 
Nucl.\ Phys.\ {\bf B191}, (1981) 429.
%%CITATION = NUPHA,B191,429;%%
\bibitem{dkmadhm} 
M. Shifman and A. Vainshtein,
``{\em Instantons versus supersymmetry: Fifteen years later}'', 
ep-th/9902018. \\ 
%%CITATION = HEP-TH 9902018;%%
N. Dorey, V. Khoze and M. Mattis ``{\em Multi-Instanton 
Calculus in $N=2$ Supersymmetric Gauge Theory}'', 
Phys. Rev. {\bf D54} (1996) 2921, hep-th/9603136. \\
%%CITATION = HEP-TH 9603136;%% 
N. Dorey, T. Hollowood and V. Khoze, ``{\em 
A Brief History of the Stringy Instanton}'', hep-th/0010015. \\
%%CITATION = HEP-TH 0010015;%%
N. Dorey, T. Hollowood, V. Khoze and M. Mattis, 
``{\em The Calculus of Many Instantons}'', To Appear.
\bibitem{vega} H. de Vega and F. Schaposnik, "{\em Classical Vortex 
Solution of the Abelian Higgs Model}", Phys. Rev. {\bf D14} (1976) 1100.
%%CITATION = PHRVA,D14,1100;%%
\bibitem{speight} J. M. Speight, ``{\em Static intervortex forces}'', 
Phys.\ Rev.\ D {\bf 55}, 3830 (1997), hep-th/9603155.
%%CITATION = HEP-TH 9603155;%%
\bibitem{dkm} N. Dorey, V. Khoze and M. Mattis, 
``{\em Multi-Instantons, Three-Dimensional Gauge Theory, 
and the Gauss-Bonnet-Chern Theorem}'', Nucl. Phys. {\bf B502} 
(1997) 94, hep-th/9704197.
%%CITATION = HEP-TH 9704197;%%
\bibitem{berg} E. Bergshoeff, B. Janssen and T. Ortin, 
"{\em Kaluza-Klein Monopoles and Gauged Sigma-Models}", 
Phys.\ Lett.\ {\bf B410}, (1997) 131, hep-th/9706117. \\
%%CITATION = HEP-TH 9706117;%%
E.~Eyras, B.~Janssen and Y.~Lozano, ``{\em 5-branes, KK-monopoles and 
T-duality}'', Nucl.\ Phys.\ {\bf B531}, (1998) 275, hep-th/9806169.
%%CITATION = HEP-TH 9806169;%%
\bibitem{ds} K. Intriligator and N. Seiberg,
``{\em Mirror symmetry in three dimensional gauge theories}'', 
Phys.\ Lett.\ {\bf B387}, (1996) 513, hep-th/9607207. \\
%%CITATION = HEP-TH 9607207;%%
D-E. Diaconescu and N. Seiberg, ``{\em The Coulomb 
Branch of (4,4) Supersymmetric Field Theories in Two Dimensions}'', 
JHEP {\bf 9707} (1997) 001, hep-th/9707158. \\
%%CITATION = HEP-TH 9707158;%%
M. Aganagic, K. Hori, A. Karch and D. Tong, 
``{\em Mirror Symmetry in 2+1 and 1+1 Dimensions}'', 
JHEP {\bf 0107} (2001) 022, hep-th/0105075.
%%CITATION = HEP-TH 0105075;%%
\bibitem{kraan} T. Kraan and P. van Baal, "{\em Exact T-Duality between 
Calorons and Taub-NUT Spaces}", 
Phys.\ Lett.\ {\bf B428}, (1998) 268, hep-th/9802049.
%%CITATION = HEP-TH 9802049;%%
\bibitem{karch} A. Karch, D. L\"ust and D. Smith, "{\em Equivalence 
of Geometric Engineering and Hanany-Witten via Fractional Branes}", 
Nucl.\ Phys.\ {\bf B533}, (1998) 348, hep-th/9803232. \\
%%CITATION = HEP-TH 9803232;%% 
B. Andreas, G. Curio and D. L\"ust, "{\em The 
Neveu-Schwarz Five-Brane and its Dual Geometries}", 
JHEP {\bf 9810}, (1998) 022, hep-th/9807008.
%%CITATION = HEP-TH 9807008;%%
\bibitem{sigmatorsion} S.~J.~Gates, C.~M.~Hull and M.~Rocek, 
``{\em Twisted Multiplets And New Supersymmetric Nonlinear Sigma Models}'', 
Nucl.\ Phys.\ {\bf B248}, (1984) 157\\
%%CITATION = NUPHA,B248,157;%%
P.~S.~Howe and G.~Sierra, 
``{\em Two-Dimensional Supersymmetric Nonlinear Sigma Models With 
Torsion}'', 
Phys.\ Lett.\ {\bf B148}, (1984) 451.\\
%%CITATION = PHLTA,B148,451;%%
T.~L.~Curtright and C.~K.~Zachos, ``{\em Geometry, Topology And 
Supersymmetry In Nonlinear Models}'', 
Phys.\ Rev.\ Lett.\  {\bf 53}, (1984) 1799.
%%CITATION = PRLTA,53,1799;%%
\bibitem{braden} H. Braden, ``{\em Sigma Models With Torsion}'', 
Annals Phys.\  {\bf 171}, (1986) 433.
%%CITATION = APNYA,171,433;%%
\bibitem{gr} G. Gibbons and P. Rychenkova ``{\em HyperKahler Quotient 
Construction of BPS Monopole Moduli Spaces}'', hep-th/9608085.
%%CITATION = HEP-TH 9608085;%%
\bibitem{ks} A. Kapustin and M. Strassler, ``{\em 
 On Mirror Symmetry in Three Dimensional Abelian Gauge Theories}'', 
JHEP {\bf 9904} (1999) 021, hep-th/9902033.
%%CITATION = HEP-TH 9902033;%%
\bibitem{rv} M. Rocek and E. Verlinde, ``{\em Duality, 
Quotients, and Currents}'', Nucl. Phys. {\bf B373} (1992) 630, 
hep-th/9110053. 
%%CITATION = HEP-TH 9110053;%%
\bibitem{kinky} N. Lambert and D. Tong, ``{\em Kinky D-Strings}'', 
Nucl. Phys. {\bf B569} (2000) 606, hep-th/9907098. 
%%CITATION = HEP-TH 9907098;%%
\bibitem{clifford} N. Mermin and H. Wagner,
``{\em Absence Of Ferromagnetism Or Antiferromagnetism In 
One-Dimensional Or Two-Dimensional Isotropic Heisenberg Models}'', 
Phys.\ Rev.\ Lett.\  {\bf 17} (1966) 1133. \\
%%CITATION = PRLTA,17,1133;%%
S. Coleman, ``{\em There Are No Goldstone Bosons In 
Two-Dimensions}'', 
Commun.\ Math.\ Phys.\  {\bf 31} (1973) 259.
%%CITATION = CMPHA,31,259;%%
\bibitem{schroers} B. Schroers, ``{\em The Spectrum of Bogomol'nyi 
Solitons in Gauged Linear Sigma Models}'', Nucl.\ Phys.\ {\bf B475}, 
(1996) 440, hep-th/9603101.
%%CITATION = HEP-TH 9603101;
\bibitem{penin} A. Penin, V. Rubakov, P. Tinyakov and S. Troitsky, 
"{\em What Becomes of Vortices in Theories with Flat Directions}", 
Phys.\ Lett.\ {\bf B389},  (1996) 13, hep-ph/9609257. \\
%%CITATION = HEP-PH 9609257;%%
A. Ach\'ucarro, A. Davis, M. Pickles and J. Urrestilla, 
"{\em Vortices in Theories with Flat Directions}", hep-th/0109097.
%%CITATION = HEP-TH 0109097;%%
\bibitem{no} H. Nielsen and P. Olesen,
``{\em Vortex-Line Models For Dual Strings}'', 
Nucl.\ Phys.\ {\bf B61} (1973) 45.
%%CITATION = NUPHA,B61,45;%%
\bibitem{erick} E. Weinberg, "{\em Multivortex Solutions of the 
Ginzburg-Landau Equations }", Phys. Rev. {\bf D19} (1979) 3008; 
"{\em Index Calculations for the Fermion-Vortex System}", 
Phys. Rev. {\bf D24} (1981) 2669.
%%CITATION = PHRVA,D24,2669;%%
%%CITATION = PHRVA,D19,3008;%%
\bibitem{taubes} C. Taubes, ``{\em Arbitrary N-Vortex Solutions To 
The First Order Landau-Ginzburg Equations}'', Commun.\ Math.\ Phys.\  
{\bf 72}, (1980), 277.
%%CITATION = CMPHA,72,277;%%
\bibitem{samols} T. Samols, ``{\em Vortex Scattering}'', 
Commun.\ Math.\ Phys.\  {\bf 145}, 149 (1992).
%%CITATION = CMPHA,145,149;%%
\bibitem{bernard} C. Bernard, ``{\em Gauge Zero Modes, Instanton 
Determinants, And Quantum Chromodynamic  Calculations}'', 
Phys. Rev. {\bf D19} (1979) 3013.
%%CITATION = PHRVA,D19,3013;%%
\bibitem{stefan} S. Vandoren and P. van Nieuwenhuizen, 
``{\em New instantons in the double-well potential}'', 
Phys.\ Lett.\ {\bf B499}, (2001) 280, hep-th/0010130.
%%CITATION = HEP-TH 0010130;%%
\bibitem{dan} D. Freedman and P. Townsend, ``{\em Antisymmetric Tensor 
Gauge Theories and Non-Linear $\sigma$-Models}''  Nucl.\ Phys. 
{\bf B177} (1981) 282. \\
%%CITATION = NUPHA,B177,282;%%
L. Alvarez-Gaum\'e and D. Freedman, ``{\em Kahler Geometry and the 
Renormalization of Supersymmetric Sigma Models}'', 
Phys.\ Rev.\ {\bf D22} (1980) 846.
%%CITATION = PHRVA,D22,846;%%
\bibitem{thooft} G.~'t Hooft, ``{\em Computation of the Quantum Effects 
due to A Four-Dimensional  Pseudoparticle}'', 
Phys.\ Rev.\ {\bf D14}, (1976) 3432. (Erratum-ibid.\ {\bf D18}, (1976))
%%CITATION = PHRVA,D14,3432;%%
\bibitem{early} N. Dorey, V. Khoze, M. Mattis, S. Vandoren and D. Tong, 
``{\em Instantons, Three-Dimensional Gauge Theory, and the Atiyah-Hitchin  
Manifold}'', Nucl.\ Phys.\ {\bf B502}, (1997) 59, hep-th/9703228. \\
%%CITATION = HEP-TH 9703228;%%
N. Dorey, S. Vandoren and D. Tong, ``{\em Instanton Effects in 
Three-Dimensional Supersymmetric Gauge Theories  with Matter}'', 
JHEP {\bf 9804} (1998) 005, hep-th/9803065.
%%CITATION = HEP-TH 9803065;%%
\bibitem{nickmartin} N. Manton and J. M. Speight,
``{\em Asymptotic Interactions of Critically Coupled Vortices}'', 
hep-th/0205307.
%%CITATION = HEP-TH 0205307;%%
\bibitem{aps} P. Argyres, R. Plesser and N. Seiberg, 
``{\em The Moduli Space of N=2 SUSY QCD and Duality in N=1 SUSY QCD}'', 
Nucl. Phys. {\bf B471} (1996) 159,  hep-th/9603042.
%%CITATION = HEP-TH 9603042;%%
\bibitem{ss} G. Segal and A. Selby, 
``{\em The Cohomology Of The Space Of Magnetic Monopoles}'', 
Commun.\ Math.\ Phys.\  {\bf 177}, (1996) 775. \\
S.~Paban, S.~Sethi and M.~Stern, ``{\em Summing up Instantons in 
Three-Dimensional Yang-Mills Theories}'', 
Adv.\ Theor.\ Math.\ Phys.\  {\bf 3}, (1999) 343, hep-th/9808119.
%%CITATION = HEP-TH 9808119;%%
%%CITATION = CMPHA,177,775;%%
\bibitem{another} N. Dorey, T. Hollowood and V. Khoze, ``{\em Notes on 
Soliton Bound-State 
Problems in Gauge Theory and String Theory}'', hep-th/0105090.
%%CITATION = HEP-TH 0105090;%%
\bibitem{dm} M. Douglas and G. Moore, ``{\em D-branes, Quivers, 
and ALE Instantons}'', hep-th/9603167.
%%CITATION = HEP-TH 9603167;%%








\end{thebibliography}
\end{document}